\documentclass[runningheads]{llncs}

\usepackage{graphicx}
\usepackage{algorithm}
\usepackage[export]{adjustbox}

\usepackage{array,xcolor,colortbl}
\usepackage{multirow}
\usepackage[outdir=./]{epstopdf}
\usepackage[export]{adjustbox}
\usepackage{siunitx}
\usepackage{array}
\usepackage{subfig}
\usepackage{color, colortbl}
\usepackage[section]{placeins}
\usepackage{newunicodechar}
\usepackage[utf8]{inputenc}
\usepackage{textcomp}
\usepackage{xcolor}
\usepackage{algorithmic}
\usepackage{pdfpages}
\usepackage{float}
\usepackage{placeins}
\usepackage{comment}
\usepackage{hyperref}
\usepackage{booktabs}

\begin{document}
\title{An Extended Variational Mode
Decomposition Algorithm Developed Speech Emotion Recognition
Performance} 

\author{David Hason Rudd\inst{1} \and
Huan Huo\inst{1,*} \and
Guandong Xu\inst{1,2,*}}
\authorrunning{D. Hason Rudd et al.}
%
\institute{The University of Technology Sydney, 15 Broadway, Ultimo, Australia \and
Data Science Institute, 15 Broadway, Ultimo, Australia\\
\email{\{david.hasonrudd, huan.huo, guandong.xu\}@uts.edu.au}\\$*$corresponding authors}

\maketitle 
\begin{abstract}

Emotion recognition (ER) from speech signals is a robust approach since it cannot be imitated like facial expression or text based sentiment analysis. Valuable information underlying the emotions are significant for human-computer interactions enabling intelligent machines to interact with sensitivity in the real world. Previous ER studies through speech signal processing have focused exclusively on associations between different signal mode decomposition methods and hidden informative features. However, improper decomposition parameter selections lead to informative signal component losses due to mode duplicating and mixing. In contrast, the current study proposes VGG-optiVMD, an empowered variational mode decomposition algorithm, to distinguish meaningful speech features and automatically select the number of decomposed modes and optimum balancing parameter for the data fidelity constraint by assessing their effects on the VGG16 flattening output layer. Various feature vectors were employed to train the VGG16 network on different databases and assess VGG-optiVMD reproducibility and reliability. One, two, and three-dimensional feature vectors were constructed by concatenating Mel-frequency cepstral coefficients, Chromagram, Mel spectrograms, Tonnetz diagrams, and spectral centroids. Results confirmed a synergistic relationship between the fine-tuning of the signal sample rate and decomposition parameters with classification accuracy, achieving state-of-the-art 96.09\% accuracy in predicting seven emotions on the Berlin EMO-DB database.

\end{abstract}
\section{Introduction}
Word meaning is often conveyed by the tone of voice~\cite{pierre2003production}, although human emotions are not solely conveyed through the words used, but also through by modifying facial expressions and vocal tone. Thus, changing voice characteristics is how most humans express different emotions~\cite{wang2010emotion}. Consequently, considerable human-computer interaction research has considered emotion recognition (ER). Various applications detect serious state by analyzing caller emotion in emergency centers; and speech pathology, e-learning, voiceprints, security, and other smart-centric services commonly employ speech emotion recognition (SER). Other approaches have considered biosensing, Electroencephalography (EEG), and facial recognition, to detect emotions~\cite{alshamsi2018automated,khare2020evolutionary,dendukuri2022emotional}.

Signal based ER employs various signals, including electrodermal activity, blood volume pulse, galvanic skin response, electrocardiogram (ECG), EEG, and speech, are commonly categorized into several decomposed modes due to the complexity and nonstationary nature of them, which allows latent factors and patterns to be extracted more easily. Several time series analysis approaches for SER have been over the previous two decades, extracting relevant speech features from nonstationary and instantaneous signals, including traditional short time Fourier transform (STFTs), empirical wavelet transforms (EWTs)~\cite{gilles2013empirical}, and variational mode decomposition (VMD)~\cite{dragomiretskiy2013variational}. Nonstationary signal properties and its components make mean STFTs are not always suitable, and previous studies have mostly considered these approaches in isolation~\cite{carvalho2020evaluating}. Huang et al.~\cite{huang1998empirical} proposed empirical mode decomposition (EMD), which decomposes the source signal into an unknown number of signal modes defined by frequency and amplitude modulated components. However, EMD has several limitations, including overlapping intrinsic mode functions (IMFs); and increased computational load when analyzing a large number of modes, particularly EEG and speech signals~\cite{carvalho2020evaluating}. 

Empirical wave transforms employ  an adaptive wavelet subdivision scheme, similar to EMD, to address EMD drawbacks by decomposing the signal into a predetermined number of IMFs or modes. Several studies have proposed an envelope weighted transformation to decompose and denoise EEG and speech signals for processing~\cite{bhattacharyya2018novel,saxena2017classification}.
Variational mode decomposition employs non-recursive decomposition to deal with nonlinear and nonstationary signals. In contrast with EWT and EMD, few studies have considered VMD to analyze EEG signals. VMD decomposes signals into modes with a narrowband around a center frequency and can overcome EWT limitations, including shift and filter bank boundary sensitivity and EMD mode mixing effects. Therefore, we were motivated to apply VMD for speech signal processing.

Acoustic feature selection is essential for SER to describe various voice signal aspects captured from different features~\cite{basharirad2017speech,dendukuri2022emotional,khare2020evolutionary}. Acoustic features include time-frequency, time, and frequency domain representations. Extracted features from time-frequency domains carry more informative data than the other domains, and better capture latent emotion content from speech signals~\cite{rudd2022leveraged}. Useful time-domain features include amplitude envelope, RMS energy, and zero-crossing rate; and are commonly employed as sequence evaluation ratios. In contrast, relevant frequency domain features include band energy ratio (BER), spectral centroid, and spectral flux. Several previous studies used VMD method to analyze signals, extracting features from the decomposed signals. However, we propose VGG-optiVMD, utilizing a VMD based feature augmentation method to enrich predictors and maximize emotion classification accuracy. Results from the proposed VGG-optiVMD approach on several common publicly available databases confirm significant ER improvement compared with previous approaches.

The main contributions from this study can be summarized as follows.
\begin{itemize}
    \item To our best knowledge, this study is the first to employ VMD as a dynamic data augmentation for speech emotion recognition.
    \item The proposed VGG-optiVMD algorithm automatically selects optimum decomposition parameters for VMD. 
    \item A reliable emotion classifier was achieved with state-of-art result 96.09\% test accuracy. 
\end{itemize}

\section{Related works}
Various approaches have been employed to detect emotions using signal processing, including analyzing EEG signals, facial expressions, speech signals, etc. Feature extraction and classification are essential but challenging to obtain optimum model performance. Most previous studies extract statistical features from the signal time-frequency domain, e.g. EEG and speech signal spectrograms, since the time-frequency domain includes more informative than the time or frequency domains for discrete-time series signals.

Dendukuri et al.~\cite{dendukuri2022emotional} decomposed the speech signal into three components sampling at 16000~Hz over 20 ms frames, then input various mode central frequency statistical parameters to a support vector machine (SVM) classifier. Optimum recognition rate achieved 85.81\% and 69.13\% accuracy for two and four emotion classes, respectively, and increased accuracy by 5\% for eight emotion classes compared to previous studies on the RAVDESS database~\cite{dendukuri2022emotional}. 

Lal et al.~\cite{lal2018epoch} empirically demonstrated VMD advantages to decompose speech signals in the correct central frequency and subsequently estimated epoch locations from noise degraded emotional speech signal.

Zhang et al.~\cite{zhang2020novel} proposed multidimensional feature extraction for EEG signal emotion recognition combining wavelet packet decomposition (WPD) with VMD to break down an EEG signals and extract wavelet packet entropy, modified multi-scale sample
entropy, fractal dimension, and first difference of each emotional  variational
mode functions as feature components. They subsequently demonstrated robust results using a random forest (RF) classifier on the DEAP dataset~\cite{koelstra2011deap}.
 
Khare et al.~\cite{khare2020evolutionary} reduced reconstruction error using meta-heuristic techniques to condensing from 16 to 1 dimension using eigenvector centrality method channel selection on EEG signals. They subsequently improved Optimized variational mode
decomposition (O-VMD) accuracy by 5\% compared with traditional VMD on the dataset of four emotions that built by themselves, with low computational load and model complexity. Furthermore, the SVM classifier significantly reducing average mean square error. 

Generally, EEG signals can effectively analyze individual's emotion since they are subject dependent. Pandey~\cite{pandey2019subject} proposed subject-independent emotion recognition using VMD and deep neural networks (VMD-DNN) on the benchmark DEAP dataset. Two features, first difference and power-spectral-density used since were sufficient to recognize calm, happy, sad, and angry emotions. SVM and DNN classifier accuracy was improved by employing VMD based feature extraction compared with EMD, STFT, and differential entropy feature extraction, achieving 61.25\% for arousal and 62.50\% valence prediction accuracies.

Several previous studies considered STFT signal decomposition techniques for SER. For example, Zhao et.~\cite{zhao2019speech} achieved robust 91.89\% accuracy on the EMODB database ~\cite{burkhardt2005database}.
Few previous studies considered VMD to decompose speech signals as mostly employed EEG signal for ER. Dendukuri~\cite{dendukuri2022emotional} achieved 69.13\% accuracy to recognize four emotions on the RAVDESS database. However, to the best of our knowledge, the current study is the first to employ VMD to enrich multidimensional feature vectors to enhance VGG-16 network learning.

\section{Proposed Methodology}
Speech signal processing involves decoding and encoding information within the speech signal. Glottal airflow from vocal folds, nasal cavity, and vocal tracts generate sounds and  words that also convey emotions. Thus, human voice is a convolution of vocal tract frequency response with a glottal pulse. The glottal pulse itself does not contain emotion related informative, and hence is considered noise in this context. The main aim for decomposition based speech signal processing is to constrain noise and interference frequencies to enhance signal decoding.

\subsection{Speech feature extraction} 
Essential and informative acoustic features in the time-frequency domain include the Mel spectrogram, chromograms, spectral contrasts, tonnetz, and Mel-frequency cepstral coefficients (MFCCs)~\cite{aizawa2004advances,harte2006detecting}. The above features are extracted and subsequently employed in various combinations to generate multidimensional feature vectors or maps.

\subsection{Variational mode decomposition}
Variational mode decomposition is a popular technique for decomposing non-stationary signals into sub-signals or modes, where mode contains a specific meaningful property from the original signal in a narrow bandwidth around the center frequency. Modes are obtained from Hilbert transform output, also called the intrinsic mode function (IMF). Furthermore, mode center frequency can be considered as a real component of the original signal for sufficiently narrow bandwidth~\cite{lal2018epoch}. The VMD adaptive algorithm reduces the original signal complexity~\cite{deb2017fourier,dragomiretskiy2013variational}.

The VMD algorithm applies the Wiener filter, Hilbert transform, analytical signals, and frequency mixing.  Wiener filters are narrowband filters for noise reduction. The Hilbert transform is a time-invariant multiplier, convolving the original signal g(t) with the impulse response $1/\pi{t}$~\cite{kschischang2006hilbert}. Therefore, it converts the real signal into the complex or imaginary part to extract magnitude and phase angle time series for frequencies with the most power at each specific time point. The VMD algorithm adds the Hilbert transform $\mathcal{H}[g(t)]$ to the original signal $g(t)$, removing any negative frequencies present (due to Hermitian symmetry). The two main VMD objects are to constrain the bandwidth for each IMF center frequency and reconstruct the original signal from the sum of all modes. First, the Hilbert transform filters frequencies in the negative side of the spectrum, and then shifts the obtained bandwidth to the modes central frequency.  Second, the obtained spectrum is shifted to the baseband region via a modulator function to obtain bandwidth around central frequency $\omega$. Finally, H1 Gaussian smoothness for the demodulation signal is used to estimate the bandwidth. Thus, constraining the L2 norm squared gradient~\cite{dragomiretskiy2013variational} defines the optimization problem (\ref{equ_min}),

\begin{equation}
\label{equ_min}
\begin{array}{r}
\min _{\left\{u_{k}\right\},\left\{\omega_{k}\right\}}\left\{\sum_{k=1}^{K}\left\|\frac{\partial}{\partial_{t}}\left[\left(\delta(t)+\frac{j}{\pi t}\right) * g_{k}(t)\right] e^{-j \omega_{k} t}\right\|_{2}^{2}\right\}  ,   \\
\text{subject to: } \quad \sum_{k=1}^{K} g_{k}(t)=g(t),  
\end{array}
\end{equation}
where the partial derivative $ \frac{\partial}{\partial_{t}} [.] $ minimizes variation in the obtained bandwidth; $ g(t)$ is the original speech signal frame; $g_{k}(t)$ is the $k{th}$ mode for $g(t)$; $K$ is the total number of modes; $\omega_{k} = \{w1, \dots ,wk\}$ is the mode center frequency, and a convenient way to reference the center frequencies for the set of $K$ modes; $e^{-j \omega_{k} t}$ is a modulator function to shift the spectrum for each mode to the baseband. 

The analytical signal generated by applying the Hilbert transform $\frac{j}{\pi t}$ and unit impulse function $\delta(t)$ as shown in equation (\ref{equ_min}). The $\delta(t)$ denotes to the Dirac delta distribution known as a unit impulse so that its value is zero everywhere and infinite at original signal. 
The original voice signal can be reproduced by solving the constraint optimization (\ref{equ_min}), which can be simplified using an augmented Lagrangian multiplier to transform it into an unconstrained problem (\ref{Lagran}), 
\begin{equation}
\begin{array}{r}
\label{Lagran}
\mathcal{L}\left(g_{k}, \omega_{k}, \lambda\right):=\alpha \sum_{k=1}^{K}\left\|\frac{\partial}{\partial_{t}}\left[\left(\left(\delta(t)+\frac{j}{\pi t}\right) * g_{k}(t)\right) e^{-j \omega_{k} t}\right]\right\|^{2} \\

+\left\|g(t)-\sum_{k=1}^{K} g_{k}(t)\right\|_{2}^{2}+\left\langle\lambda{(t)}, g{(t)}-\sum_{k=1}^{K} g_{k}(t)\right\rangle,   
\end{array}
\end{equation}

where, $\lambda$ is a time dependent Lagrangian multiplier, and $\alpha$ is a bandwidth control parameter. 

The unconstrained Lagrangian problem (\ref{Lagran}) can be solved to obtain the frequency and the modes using the alternate direction method of multipliers (ADMM)~\cite{hestenes1969multiplier,rockafellar1973dual,dragomiretskiy2013variational} optimization in spectral domain. However, optimization outcomes are the same for the frequency and time domains Hence mode $g_{k}(\omega)$ can be updated in the spectral domain,
\begin{equation}
\label{gk}
\hat{g}_{k}^{n+1}(\omega) \leftarrow \frac{\hat{g}(\omega)-\sum_{i<k} \hat{g}_{i}^{n+1}(\omega)-\sum_{i>k} \hat{g}_{i}^{n}(\omega)+\frac{\hat{\lambda}^{n}(\omega)}{2}}{1+2 \alpha\left(\omega-\omega_{k}^{n}\right)^{2}}  .  
\end{equation}

Updating is obtained using the Wiener filter for the current residual using the signal prior $1 / (\omega - \omega_{k})^2$ to restrain variation across the central frequency minimum, providing the updated mode center frequency $\omega_{k}$ as
\begin{equation}
\label{wk)}
\hat{\omega}_{k}^{n+1}=\frac{\int_{0}^{\infty} \omega\left|\hat{G}_{k}(\omega)\right|^{2} \mathrm{~d} \omega}{\int_{0}^{\infty}\left|\hat{G}_{k}(\omega)\right|^{2} \mathrm{~d} \omega},  
\end{equation}
where $\hat{G}_{k}(\omega)$ is the Fourier transform for $g_{k}^{n+1}(t)$. 

A better decomposed signal can be obtained by reconstructing the original signal as the sum of modes and estimating bandwidth using the Wiener filter. Details for the VMD algorithm are provided in~\cite{dragomiretskiy2013variational}.

To leverage VMD effectiveness, we propose the VGG-optiVMD algorithm for automatically selecting optimum $\alpha$ and $K$ by analyzing different decomposition parameter effects on classification accuracy.  

\subsection{Proposed VGG-optiVMD}
Reconstruction error for a decomposed signal can be reduced by selecting optimum $K$ and $\alpha$. Improper decomposition parameter selection will create duplicate modes, causing signal information losses and hence reduced classifier performance. 
One drawback for VMD is that finding decomposition parameters $K$ and $\alpha$ to provide optimum performance challenging. Several approaches have proposed for ER using ECG, EEG and vibrational signals. For example, the OVMD algorithm ~\cite{lian2018adaptive} uses a series of indicators, including permutation entropy, kurtosis criteria, extreme frequency domain value, and energy loss coefficients, to identify optimum $K$. Wang et al.~\cite{wang2019application} controlled power spectral and dynamic entropy features to find optimal $K$ and $\alpha$ to decompose vibration signal and extract fault features.  

However, these approaches use IMF or mode characteristics to find the best decomposition parameters for specific low amplitude input signals with empirical threshold selection, which is not applicable for speech signal processing. Dendukuri et al.~\cite{dendukuri2022emotional} decomposed speech signals using five modes to recognize eight emotions, achieving 61.2\% accuracy on the RAVDESS database. They combined different features, including a 45-dimensional feature set including mode center frequency, statistical values for mode center frequency, MFCCs, and spectral statistical features to improve classifier performance.

The above methods evaluate optimum $K$ value using statistical features and indicators for guidance. In particular, identified mode number correctness was not verified or fine-tuned practically by monitoring classification accuracy.

In contrast, the current study proposes to automate optimum VMD decomposition parameter selection using a feedback loop from the VGG16 flattening output layer. Algorithm 1 shows the proposed optimized VMD algorithm (VGG-optiVMD). The key strength for VGG-optiVMD is reliability, generality, and reproducibility across different speech databases for real-world applications, e.g. customer satisfaction analysis in call centers.

\subsection{Feature scaling, data augmentation, and emotion classification} 
\begin{figure}[h]
    \centering  
        \includegraphics[scale=0.14]{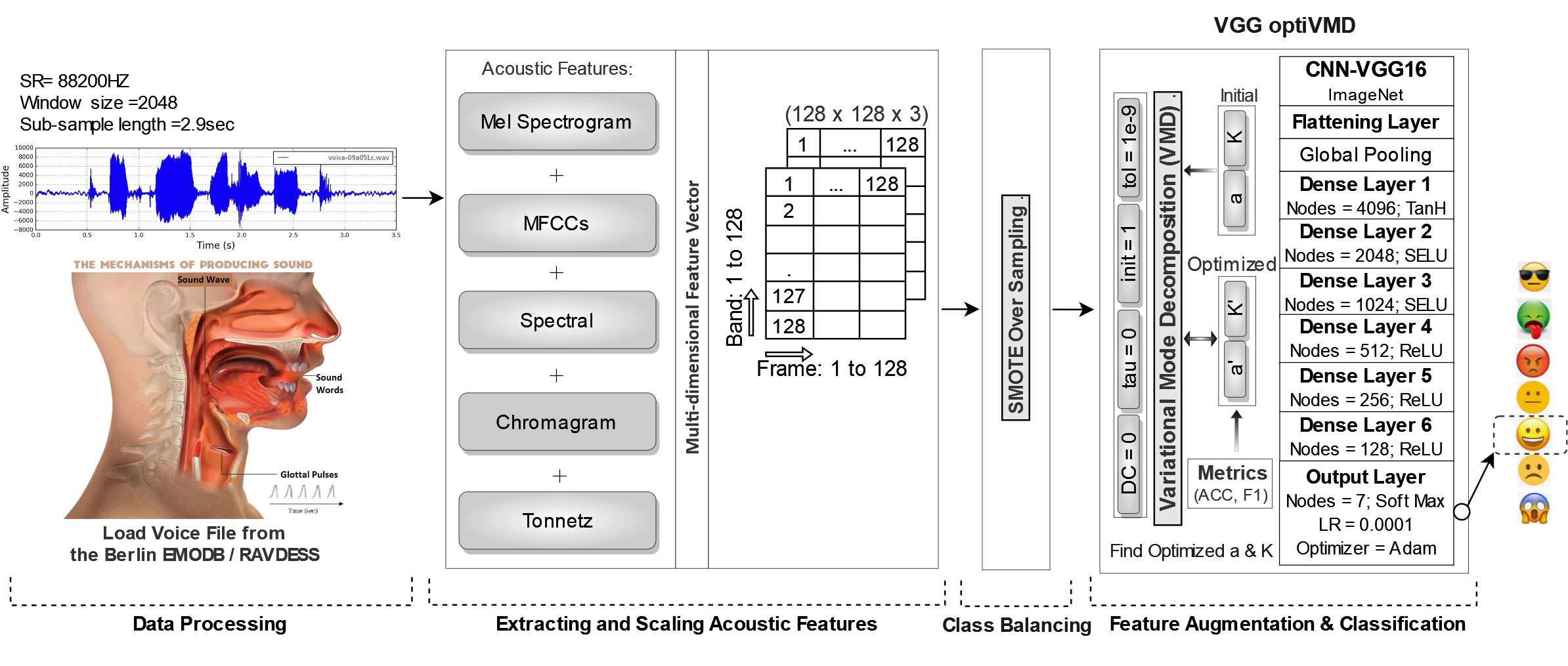}
    \caption{Proposed model development workflow: extracted features are enriched using the VGG-optiVMD to automatically identify $K$ and $\alpha$.}
    \label{FIG:1}
\end{figure}

Figure \ref{FIG:1} shows the proposed framework to train CNN-VGG16~\cite{russakovsky2015imagenet} to extract enriched feature vectors and classify seven emotions: anger, boredom, happiness, neutral, disgust, sadness, and fear on two databases EMODB and RAVDESS. 
Figure \ref{FIG:1} shows the model development proceeds as follows.
\begin{enumerate}
\item The voice signal is sampled at 88400 Hz and five well-known acoustic features extracted in the time-frequency domain: MFCCs, Mel spectrogram, Tonnetz, spectral contrast, and chromagram. 
\item The Hann window function is applied with 2.9 s fixed length and 0.4~ms shifting time to sub-signal spectra assembled over a series of frames, extracted features are reshaped into a single $(128 \times 128 \times 3)$ feature vector.
\item The SMOTE~\cite{livingstone2018ryerson} oversampling strategy is applied to compensate minority classes and reduce model bias. Final testing and training features are randomly partitioned into 20\% and 80\% sets, respectively.
\item The proposed VGG-optiVMD algorithm is applied to decode frequency statistical properties at specific times that distinguish emotions within the feature vector. 
\item The VGG network is trained on the augmented feature vector to classify emotions into seven classes.
\end{enumerate}

\section{Experiment Setup} 
This study followed the preprocessing system from~\cite{rudd2022leveraged}. All acoustic features were extracted using the Librosa tool~\cite{mcfee2015librosa} using the Ryerson Audio Visual Database of Emotional Speech and  Song (RAVDESS)~\cite{livingstone2018ryerson} and the Berlin EMODB~\cite{burkhardt2005database} databases. Voice data are preprocessed with frame size = 2048, HOP length = 256, and sampling rate = 88400 to avoid spectral leakage and enhance frequency resolution. Several experiments were performed on nine different feature vectors to identify the proposed VGG-optiVMD algorithm effectiveness using. The model was implemented on a Keras framework. The detail of network implementations are available in our GitHub repository\footnote{https://github.com/-----}. 

\subsection{Modelling}

The aim of modeling was to enhance informative data within the feature vectors and avoid overfitting. Therefore, we applied data augmentation by decomposing the feature vector data, i.e., $g(t)$ is explained in proposed algorithm, into different modes. Augmentation effects on classification accuracy were assessed using diverse $K$ and $\alpha$ sets.
Optimal $K$ and $\alpha$ was assessed iteratively until robust classification accuracy was achieved or the break loop condition reached. $K$ and $\alpha$ were set to a wide range of 3--8 and 1000--6000, respectively, based on empirical experiments since there was no significant improvement in prediction accuracy outside those ranges. The VGG16 is selected to be trained from augmented feature maps as a trade-off between model runtime and classifier accuracy. The VGG16 architecture used the ADAM optimizer with learning rate = 0.0001; six fully connected hidden layers with ReLU, SELU, and TanH activation functions; epochs = 50, batch size = 4; and SoftMax function for the output layer.

\section{Result and Discussion}
To assess the effectiveness of our VMD-based feature augmentation method several evaluation metrics were employed including F1 score, test accuracy, and confusion matrix. Based on the experiment results shown in Table 1, there is a correlation between the number of modes $K$, $\alpha$ and classification accuracy. The different acoustic features are enriched with various sets of decomposition parameters. Results showed that higher accuracy was obtained for $K$ (4 - 6) and $\alpha$ (2000 - 4000) in both datasets, although VGG-optiVMD is set to a limited range of $\alpha$ (1000-10000) and $K$ (2-8) due to increasing a heavy computational load when $K$ value is over 8 with sample rate 88400. This limitation can be considered a functional constraint of VGG-optiVMD. Nevertheless, a state-of-the-art result was achieved with the accuracy of 96.09\% with $K$=6 and $\alpha$=2000 as demonstrated in Table~\ref{Ka}.

\begin{figure}[!h]
    \centering
    \includegraphics[scale=0.4]{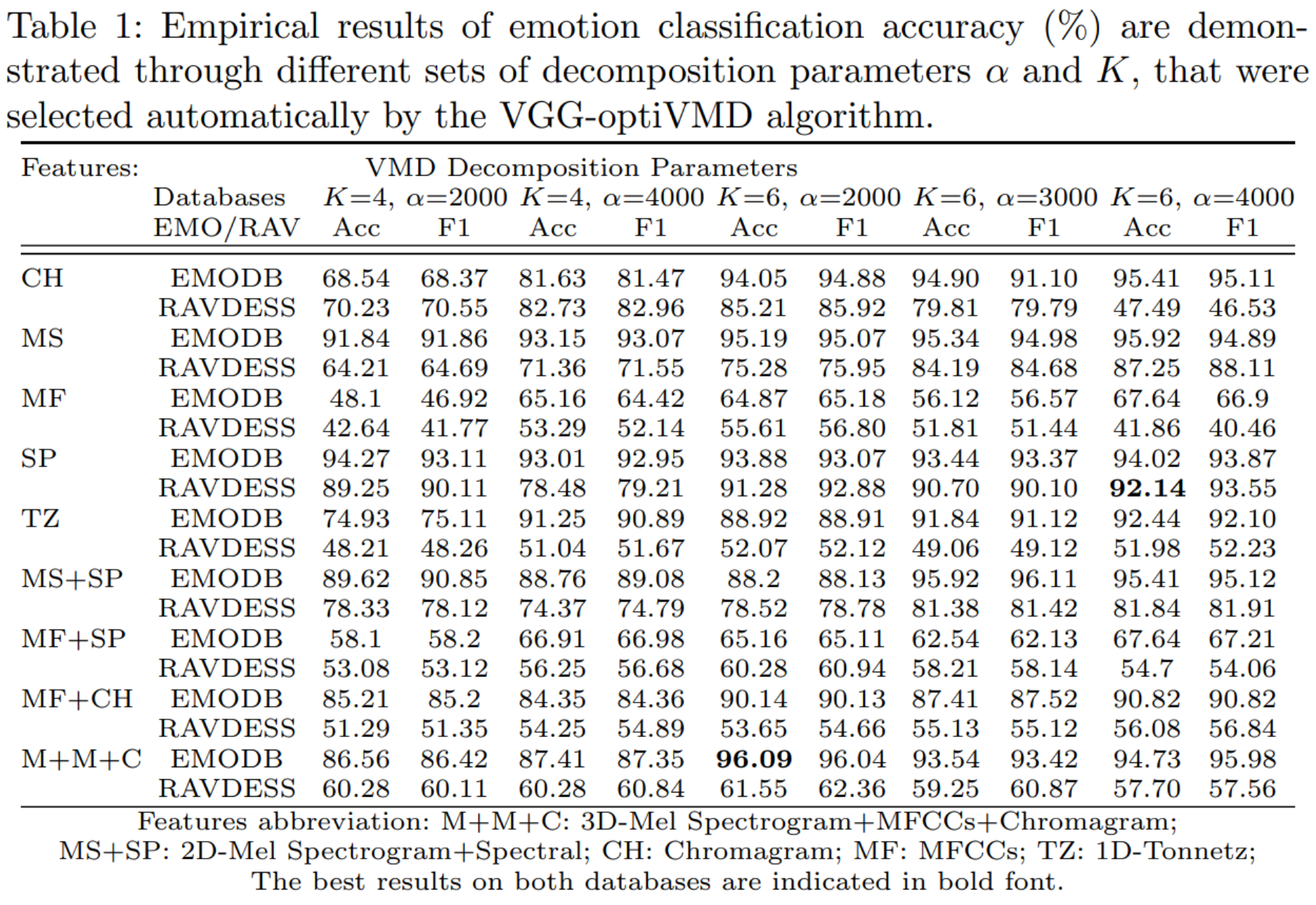}
    \caption{Empirical results of emotion classification accuracy (\%) are demonstrated through different sets of decomposition parameters $\alpha$ and $K$, that were selected automatically by the VGG-optiVMD algorithm.}
    \label{fig:enter-label}
\end{figure}

\begin{figure}[h]
	\centering
		\includegraphics[scale=0.08]{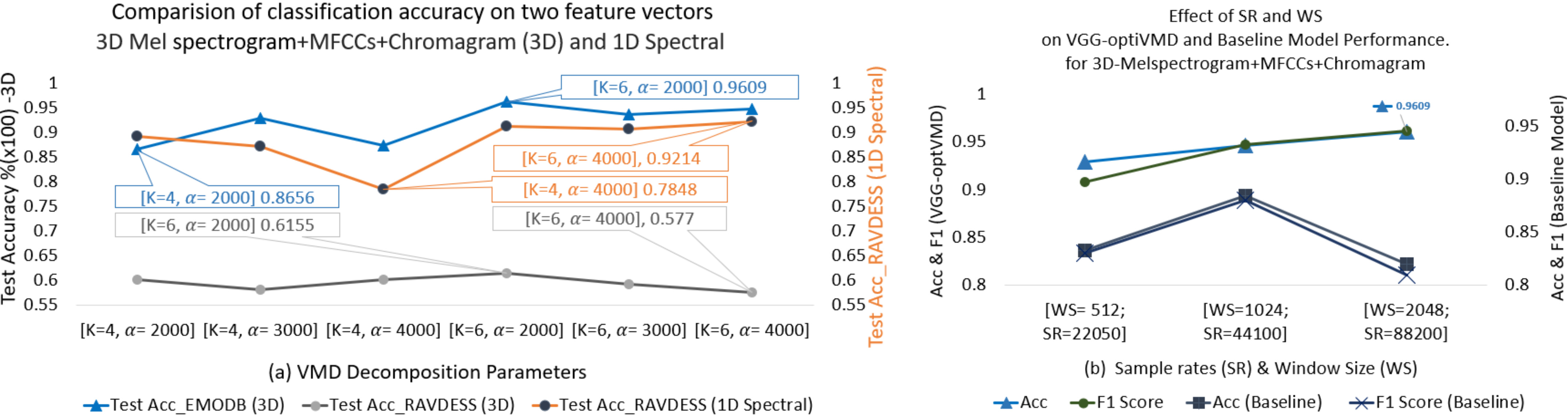}
	\caption{The model performance is assessed by different signal sampling rates and VMD parameters $K$ and $\alpha$. Graph (a) The VGG-optiVMD identified the set of $K$ = 6 and $\alpha$ = 2000 as optimum value. Graph(b) represents the effect of various ranges of sample rate and window size on the proposed and baseline model in EMODB. The highest accuracy can be achieved by SR = 88200 and WS = 2048.}
	\label{FIG:2}
\end{figure}

\begin{figure}[t]
	\centering
		\includegraphics[scale=0.087]{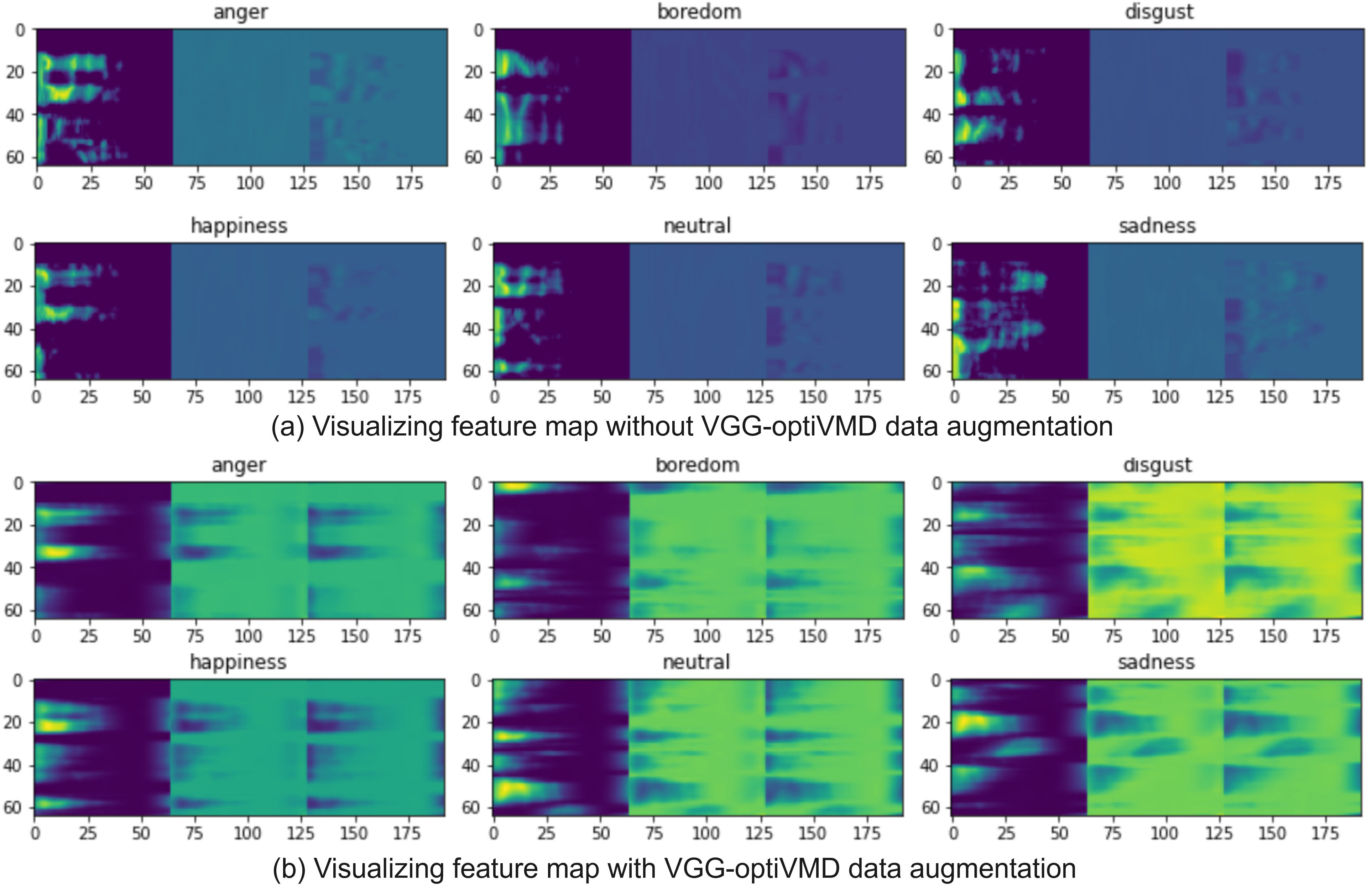}
	\caption{The efficient functionality of VGG-optiVMD on the feature vector 3D-Mel Spectrogram+MFCCs+Chromagram clearly shows a higher distinction in the energy magnitudes of frequencies in (b).}
	\label{FIG:3}
\end{figure}

Analyzing the results of the baseline model, which is built with the same framework simply without VMD-based feature vector augmentation, helps us to justify the power of the VGG-optiVMD in SER. Therefore, we attempted to evaluate the model performance through variation of sample rate, window size, $K$ and $\alpha$ without using VMD (baseline model) and with VMD (proposed model). 
As shown in Figure \ref{FIG:2}, unlike the baseline model, the proposed model performed better with a larger sampling rate and window size. Moreover, the highest test accuracy and F1 score were obtained via VGG-optiVMD, proving that our VMD-based feature augmentation method significantly improved the classification accuracy. 
The Figure \ref{FIG:3} shows the efficient functionality of VGG-optiVMD on the feature vector 3D-Mel Spectrogram+MFCCs+Chromagram. Figure (a) represents the feature before applying VMD based data augmentation, and figure (b) clearly shows that the informative frequencies are distinguished on the feature vector after applying the data augmentation method. In addition, the image shows the feature vector acquired higher distinction energies in the time-frequency domain. Therefore, the implications of this finding can improve the learning process in VGG16 and result in better prediction accuracy.
The confusion matrix in Table \ref{Confusion} demonstrates the high performance of the classification model with accuracy above 90\% for all classes. Nevertheless, the model performs poorly when predicting happiness and anger emotions due to the similarity of signal attributes such as intensity, frequency and harmonic structure.

\begin{table}[t]
\centering
\scriptsize
\caption{Visualization of the model performance with confusion matrix (\%) for the 3D-Mel Spectrogram+MFCCs+Chromagram with test accuracy = \%96.09 on the Berlin EMO-DB dataset.}
\label{tab:Datasets}
\begin{tabular}{llccccccc} 
\hline
\ \ & \ \ Emotion:  &  Anger  &  Boredom  &  Disgust  &  Fear  &  Happiness  &  Neutral &  Sadness\\ \hline

\ \ & \ \ Anger \ \ & \ \ \bfseries95.24 \ \ & \ \ 0 \ \ & \ \ 0 \ \ & \ \ 0 \ \ & \ \ 4.76 \ \ & \ \ 0 \ \ & \ \ 0 \ \

\\

\ \ & \ \ Boredom \ \ & \ \ 0 \ \ & \ \ \bfseries95.24	 \ \ & \ \ 0 \ \ & \ \ 0 \ \ & \ \ 0 \ \ & \ \ 0 \ \ & \ \ 4.76 \ \

\\

\ \ & \ \ Disgust \ \ & \ \ 0 \ \ & \ \ 0 \ \ & \ \ \bfseries100.00	 \ \ & \ \ 0 \ \ & \ \ 8 \ \ & \ \ 0 \ \ & \ \ 0 \ \

\\

\ \ &\ \ Fear \ \ & \ \ 0 \ \ & \ \ 0 \ \ & \ \ 0 \ \ & \ \ \bfseries94.05 \ \ & \ \ 0 \ \ & \ \ 0 \ \ & \ \ 0 \ \

\\

\ \ & \ \ Happiness \ \ & \ \ 8.33 \ \ & \ \ 0 \ \ & \ \ 0 \ \ & \ \ 0 \ \ & \ \ \bfseries91.67 \ \ & \ \ 0 \ \ & \ \ 0 \ \

\\

\ \ & \ \ Neutral \ \ & \ \ 0 \ \ & \ \ 2.38 \ \ & \ \ 0 \ \ & \ \ 0 \ \ & \ \ 1.19 \ \ & \ \ \bfseries96.43 \ \ & \ \ 0 \ \

\\

\ \ & \ \ Sadness \ \ & \ \ 0 \ \ & \ \ 0 \ \ & \ \ 0 \ \ & \ \ 0 \ \ & \ \ 0 \ \ & \ \ 0 \ \ & \ \ \bfseries100 \ \ \\ \hline
\end{tabular}
\label{Confusion}
\end{table}
The VGG-optiVMD method is compared with the most recent works, shown in Table \ref{Comparison}, that our method outperforms previous models and achieves a state-of-the-art result in terms of accuracy. In accordance with the knowledge we have, this is the first work to employ VMD as a feature augmentation method in SER. Moreover, the main advantage of the VGG-optiVMD is its generality, which can be employed independently for other acoustic features and different databases.
\begin{table}[!h]
\centering
\caption{Comparison of the proposed method with previous works on the same databases.} 
\label{Comparison}
\scriptsize
\begin{tabular}{llll} \hline
Method proposed by  &  Feature extraction strategy & Learning Net. &  Acc(\%) \\ \hline
Badshah et al.~\cite{badshah2017speech}   & log Mel spectrogram & CNN   & 52 \\ Dendukuri et al.~\cite{dendukuri2022emotional} & 45d- Mode statistical+MFCCs+Spectral & SVM & 61.2 \\
Zamil et al.~\cite{zamil2019emotion} & 13 MFCCs & Tree Model & 70 \\
Popova et al.~\cite{popova2017emotion}  &  Mel spectrograms & VGG16   &  71 \\ 
Hajarol. et al.~\cite{hajarolasvadi20193d}  &  Mel spectrograms+MFCCs & CNN   & 72.21 \\ 
Wang et al.~\cite{wang2015speech} &   Fourier Parameter+MFCCs & SVM  & 73.3 \\ 
Kown et al.~\cite{kwon2019cnn} & Spectrogram & Deep SCNN & 79.50 \\
Badsha et al.~\cite{badshah2019deep} & Spectrogram & CNN & 80.79  \\
Huang et al.~\cite{huang2014speech}   &  Spectrogram & CNN  & 85.2 \\ 
Issa et al.~\cite{issa2020speech}  &  MFCCs+Chroma.+Mel spec.+Contrast+Tonnetz & VGG16  & 86.10 \\ 
Meng et al.~\cite{meng2019speech}  &  log Mel spec.+1st \& 2nd delta(log Mel spec.) & CNN-LSTM  & 90.78 \\ 
Wu et al.~\cite{wu2011automatic}  &  Modulation Spectral Features (MSFs) & SVM   & 91.60 \\
Rudd et al.~\cite{rudd2022leveraged}  &   Harmonic-Percussive (HP)+log Mel spec. & VGG16-MLP  &  92.79 \\ 
Demircan et al.~\cite{demircan2018application}  &  LPC+MFCCs & SVM   & 92.86 \\ Zhao et al.~\cite{zhao2019speech}   &  log Mel spectrogram & CNN-LSTM  & 95.89 \\ 
\bfseries VGG-optiVMD  &  \bfseries 3D-Mel spectrogram+MFCCs+Chromagram  & \bfseries VGG16-VMD &  \bfseries 96.09\% \\ \hline
\end{tabular} 
\end{table}
\section {Conclusion}
Speech signal processing is employed in some applications when we only have access to speech voice to detect emotions which is the first aim of this study, the second aim of this study is to introduce specific data augmentation techniques to enrich the extracted acoustic features by design of VGG-optiVMD, an extended VMD algorithm to improve SER performance.

The findings provide solid empirical confirmation of the key role of the sampling rate, the number of the decomposed mode, $K$ and the balancing parameter of the data-fidelity constraint, $\alpha$, in the performance of the emotion classifier. Taken together, these findings suggest that VMD decomposition parameters $K$ (2-6) and $\alpha$ (2000-6000) are optimum values on both the RAVDESS and EMODB databases. 
The proposed VGG-optiVMD algorithm improved the emotion classification to a state-of-the-art result with a test accuracy of 96.09\% in the Berlin EMO-DB and 86.21\% in the RAVDESS datasets. 
Further work needs to be done to establish whether extracting acoustic features only from informative decomposed modes can reduce computational load constraints. Therefore, the study should be repeated using the VMD algorithm before acoustic feature extraction process. 
%
%

%

%
%
\bibliographystyle{splncs04}

\bibliography{mybibliography}

%






\end{document}